\documentclass[twocolumn,showpacs,showkeys,pra]{revtex4}
\usepackage{graphicx}%
\usepackage{dcolumn}
\usepackage{amsmath}
\usepackage{latexsym}

\begin{document}

\def\f{\frac}
\def\rdown{\rho_{\downarrow}}
\def\pa{\partial}
\def\th{\theta}
\def\Ga{\Gamma}
\def\ka{\kappa}
\def\bea{\begin{eqnarray}}
\def\eea{\end{eqnarray}}
\def\be{\begin{equation}}
\def\ee{\end{equation}}
\def\pa{\partial}
\def\d{\delta}
\def\K{\kappa}
\def\a{\alpha}
\def\eps{\epsilon}
\def\th{\theta}
\def\na{\nabla}
\def\nn{\nonumber}
\def\lan{\langle}
\def\ran{\rangle}
\def\pr{\prime}
\def\rarrow{\rightarrow} 
\def\larrow{\leftarrow}

\draft
\title{Phase delay time and superluminal propagation in barrier tunneling}
\author{Swarnali Bandopadhyay}
\affiliation{S. N. Bose National Centre for Basic Sciences, JD Block,
Sector III, Salt Lake City, Kolkata 700098, India}
\email{swarnali@bose.res.in} 
\author{A. M. Jayannavar}
\email{jayan@iopb.res.in}  
\affiliation{Institute of Physics, Sachivalaya Marg, 
Bhubaneswar 751005, India}

\date{\today}

\begin{abstract}                          
In this work we study the behaviour of Wigner phase delay time for
tunneling in the reflection mode. Our system consists of a circular loop
connected to a single wire of semi-infinite length in the presence of
Aharonov-Bohm flux. We calculate the analytical expression for the
saturated delay time.
This saturated delay time is independent of Aharonov- Bohm flux and the
width of the opaque barrier thereby generalizing the Hartman effect. This
effect implies superluminal group velocities as a consequence. We also
briefly discuss the
concept called ``space collapse or space destroyer".
\end{abstract}

\vskip 0.5cm 

\maketitle
\section{INTRODUCTION}
\label{s1}
Quantum tunneling is one of the most important phenomenon with
wide range of applications in modern technology. In spite of its remarkable
success in various applications, there still remains the fundamental
question, namely, how much time does a particle take to traverse the
barrier? ( so called tunneling time problem). This problem has been
approached from many different viewpoints, but still there is a lack
of consensus about the existence of simple expression for this time
\cite{landauer,hauge,recami}.
This is due to the fact that there is no Hermitian operator
associated with this time. The various time scales proposed in the
literature
are based on different operational definitions and physical
interpretations.
They may represent different complimentary aspect of tunneling process.
Some of them include dwell time, Larmor time, complex times,
Buttiker- Landauer times,
sojourn times \cite{anantha}, 
time scale based on Bohm's view \cite{jayan1},
phase delay time, etc.

In our present work we concentrate mainly on the concept of
Wigner phase delay time. This time is usually taken as the difference
between the time at which the peak of the transmitted packet leaves the
barrier and the time at which the peak of the incident Gaussian
wave packet arrives at the barrier. Within the stationary phase approximation
the phase time can be calculated from the energy derivative of the 
`phase shift' in the transmitted or reflected amplitudes. This phase delay time is also related to various physical quantities such as partial density of states \cite{buttiker}. 
In the case of a quantum tunneling it has been shown that in the opaque
barrier limit the phase delay time does not depend on the barrier
width. This phenomenon is called  as the Hartman effect.
This implies that for sufficiently large
barriers the effective velocity of the particle can become
arbitrarily large,  even larger than the  speed of light in the
vacuum (superluminal). This may be further regarded as the evidence
of the fact that quantum systems seem to behave as non-local.
Experiments have clearly demonstrated \cite{steinberg,nimtz,expt} that 
``tunneling photons"
travel with superluminal group velocities.
Their measured tunneling time is practically
obtained by comparing the two peaks of the incident and
transmitted wave packets. It is important to note that these observations 
do not violate `Einstein causality', i.e., the signal velocity or 
the information transfer velocity is always bounded by the velocity of light.
Theoretically Hartman effect has been generalized to different
cases including double barriers, various geometric structures and in
the presence of Aharonov-Bohm flux. Mystery surrounding this effect 
has also been addressed recently by Winful. He argues that the short time 
delay observed is due to energy storage and release and has nothing to do with
propagation. Since the
stored energy or the probability density in the evanescent field decreases
exponentially with the distance in the barrier, after certain decay
distance it does not matter how much more length the barrier has. For
details we refer to \cite{winful}.

\begin{figure}[t]
\begin{center}
\includegraphics[width=8.0cm]{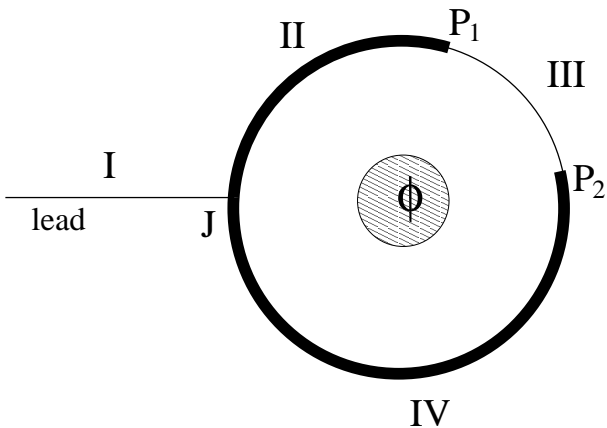}
\end{center}
\caption{Schematic diagram of a circular loop connected to semi infinite lead.} 
\label{system}
\end{figure}

    In our present work we study the above mentioned effect in a simple
quantum system consisting of a circular loop connected to a single wire
of semi-infinite length as shown in Fig~\ref{system}. In addition we impose
Aharonov-Bohm flux through the loop. We also focus on the following situation.
There is a potential
barrier (or barriers) inside the loop, while the potential in the
connecting lead is set to zero. The incident energy of the free
propagating electron $E$ in the semi-infinite wire is less than the
barrier potential height $V$. The impinging electron on the loop in
the sub barrier region travels as an evanescent wave in the loop
before being reflected. We analyze the phase time of the
reflected wave.
We show that this phase time in the opaque barrier regime
becomes independent of the length of the circumference of the ring
and the magnitude of the Aharonov-Bohm
flux, thereby generalizing the Hartman effect beyond one dimension
and in the presence of magnetic flux. We have also extended this effect by
including an additional potential well in the circular ring. Interestingly 
the saturated delay time becomes independent of the length of the 
potential well (in the large length limit) in the off resonant case.
The potential well can support many resonant states. This result is
regarded as a ``space collapse or space destroyer" \cite{recami2}. Even though in the potential well
inside the loop electronic wave can travel as a free propagating
mode (and not as a evanescent mode), surprisingly the saturated delay
time is independent of the length of the well (as if it does not count).

    In the sections to follow we first give the description
of the theoretical treatment. For this we make use of the quantum
waveguide approach well known for the electronic transport problems.
The classical relativistic Helmholtz equation for the electromagnetic wave 
propagation is formally identical to non relativistic stationary Schr\"odinger 
equation. Hence conclusions obtained using quantum tunneling theory are equally valid in the case of electromagnetic phenomenon ( photon tunneling ). In the 
later sections we will discuss our results and conclusions.

\section{THEORETICAL TREATMENT}
We approach this scattering problem using the quantum wave guide theory 
~\cite{xia,jayan3}. In the stationary case the incoming particles
are represented by a plane wave $e^{ikx}$ of unit amplitude. The
effective mass of the propagating particle is $m$ and the energy is 
$E=\hbar^2 k^2/2m$ where $k$ is the wave vector corresponding to the free 
particle.
The wave function 
in different regions (which are solutions of the Schr\"odinger equation) 
in the absence of magnetic flux are given below 
\begin{eqnarray}
\psi_0(x_0) &=& e^{ikx_0} + R e^{-ikx_0}\,\,\,\,\,\mbox{( in region I )}\label{wv1}\\
\psi_1(x_1)&=& A_1\, e^{i\,q_1\,x_1} + B_1\, e^{-i\,q_1\,x_1}\,\,\,\,\, \mbox{( in region II )} \label{wv2}\\
\psi_2(x_2) &=& A_2\, e^{i\,k\,x_2} + B_2\, e^{-i\,k\,x_2} \,\,\,\,\, \mbox{( in region III) }\label{wv3}\\
\psi_3(x_3) &=& A_3\, e^{i\,q_3\,x_3} + B_3\, e^{-i\,q_3\,x_3}\,\,\,\,\, \mbox{( in region IV) }\label{wv4}
\end{eqnarray}
with $k$ being the wavevector of electrons in the lead and in the intermediate 
free space between two barriers inside the ring. $q_1=\sqrt{2m(E-V_1)/\hbar^2}$ and $q_3=\sqrt{2m(E-V_3)/\hbar^2}$ are the wavevector respectively for propagating electrons in the barriers of strength $V_1$ and $V_3$ inside the ring. The 
origin of the co-ordinates of 
$x_0$ and $x_1$ is assumed to be at $J$ and that for 
$x_2$ and $x_3$ are at $P_1$ and $P_2$ respectively. At $P_1$, $x_1=lb_1$, at
$P_2$, $x_2=w$ and at $J$, $x_3=lb_3$, where $lb_1$ and $lb_3$ are the
length of the two barriers separated by a well region of length $w$ inside the ring. Total circumference of the ring is $L=lb_1 + lb_3 + w$.

We use Griffith's boundary conditions~\cite{griffith}
\begin{equation}
\psi_0(0)=\psi_1(0)=\psi_3(lb_3) \, \label{bc1}
\end{equation}
and 
\begin{equation}
\frac{\partial \psi_0(x_0)}{\partial x_0}\Big |_J + \frac{\partial \psi_1(x_1)}{\partial x_1}\Big |_J + \frac{\partial \psi_3(x_3)}{\partial x_3}\Big |_J
=0 \, ,
\label{bc2}
\end{equation}
at the junction $J$. All the derivatives are taken either outward or inward 
from the junction~\cite{xia}. Similar boundary conditions are held at $P_1$
and $P_2$ {\it i.e.}
\begin{eqnarray}
\mbox{at \, \, $P_1$ \,:}\,\,\,\,\,\,\,\,\,\,\,\,\,\,\psi_1(lb_1)=\psi_2(0)\\
\frac{\partial \psi_1(x_1)}{\partial x_1}\Big |_{P_1} + 
\frac{\partial \psi_2(x_2)}{\partial x_2}\Big |_{P_1}=0 \, , \\
\mbox{at \, \, $P_2$ \,:}\,\,\,\,\,\,\,\,\,\,\,\,\,\,\,\psi_1(w)=\psi_3(0)\\
\frac{\partial \psi_2(x_2)}{\partial x_1}\Big |_{P_2} + 
\frac{\partial \psi_3(x_3)}{\partial x_3}\Big |_{P_2}=0 \, .
\label{bc2}
\end{eqnarray}

We choose a gauge for the vector potential in which the magnetic field appears
only in the boundary conditions rather than explicitly in the Hamiltonian~\cite{xia,gefen}. Thus the electrons propagating clockwise and anticlockwise will pick up opposite phases. 
The electrons propagating in the clockwise direction from $J$ will pick up phases 
$i\,\a_1$ at $P_1$, $i\,(\a_1+\a_2)$ at $P_2$ and $i\,(\a_1+\a_2+\a_3)$ at $J$ after 
traversing once along the loop. The total phase around the loop is 
$\a_1\,+\,\a_2\,+\,\a_3\,=\,2\,\pi\,\phi/\phi_0$, where $\phi$ and 
$\phi_0$ are the magnetic flux and flux quantum, respectively. 
Hence from above mentioned boundary conditions we get for tunneling particle
\begin{widetext}
\begin{eqnarray}
1+R-A_1-B_1 \exp(-i\,\a_1)=0\, ,\\
A_3\,\exp(-\kappa_3\,lb_3)\,\exp(i\a_3)+B_3\,\exp(\kappa_3\,lb_3)-1-R=0\, ,\\
ik\,(1-R)+\kappa_1\,(A_1-B_1\,\exp(-i\,\a_1))\hskip 2cm \nn\\
-\,\kappa_3\,A_3\,\exp(-\kappa_3\,lb_3)\,\exp(i\,\a_3)
-\,\kappa_3\,B_3\,\exp(\kappa_3\,lb_3)=0\, ,\\
A_1\,\exp(-\kappa_1\,lb_1)\,\exp[i\,\a_1]+B_1\,\exp(\kappa_1\,lb_1)\hskip 2cm \nn\\
-\,A_2-B_2\,\exp(-i\,\a_2)=0\, ,\\
\kappa_1\,A_1\,\exp(-\kappa_1\,lb_1)\,\exp(i\,\a_1)-\kappa_1\,B_1\,\exp(\kappa_1\,lb_1)\hskip 1cm \nn\\
+\,i\,k\,A_2-i\,k\,B_2\,\exp(-i\a_2)=0\, ,\\
A_2\,\exp(i\,k\,w)\,\exp(i\,\a_2)\,+\,B_2\,\exp(-i\,k\,w)\hskip 2cm \nn\\
-\,A_3-B_3\,\exp(-i\a_3)=0\, ,\\
i\,k\,A_2\,\exp(i\,k\,w)-i\,k\,B_2\,\exp(-i\,k\,w)\,\exp(-i\a_2)\,\hskip 1cm \nn\\
+\,\kappa_3\,A3-\kappa_3\,B_3\,\exp(-i\,\a_3)=0\, ,
\end{eqnarray}
\end{widetext}
with $\kappa_1=\sqrt{2m(V_1-E)/\hbar^2}$ and 
$\kappa_3=\sqrt{2m(V_3-E)/\hbar^2}$ being the imaginary wave vectors, 
in presence of rectangular barriers of
strength $V_1$ and $V_3$ respectively, inside the ring.
\section{Results and Discussion}
\label{result}
Once $R$ is known, the `reflection phase time' $\tau$ can be calculated from 
the energy derivative of the phase of the reflection amplitude \cite{hauge,wigner} as
\begin{equation}
\tau = \hbar \, \frac{\partial Arg[R]}{\partial E}\, , \label{phtm}
\end{equation}
where, $v = \hbar\,k/m$ is the velocity of the free particle. The concept of 
`phase time' was first introduced by Wigner \cite{wigner} to estimate how long 
a quantum mechanical wave packet is delayed by the scattering obstacle.

For a ring system with a rectangular barrier of strength $V_1$ along its entire circumference we obtain an analytical expression for the reflection amplitude as
\begin{equation}
R=\frac{-\kappa_1\,(2\,\cos(\a)-\exp(kL))+i\,\frac{k}{2}\,\exp(kL)}
{\kappa_1\,(2\,\cos(\a)-\exp(kL))+i\,\frac{k}{2}\,\exp(kL)}\, ,\label{ref}
\end{equation} 
where $\a=\a_1\,+\,\a_2\,+\,\a_3$. In what follows, let us set $\hbar =1$ 
and $2m =1$. We now proceed to analyze
the behavior of $\tau$ as a function of various 
physical parameters for different ring systems. We express all the
physical quantities in dimensionless units {\em i.e.} all the barrier strengths 
$V_n$ in units of incident energy $E$ ($V_n \equiv V_n/E$), all the
barrier widths $lb_n$ in units of inverse wave vector 
$k^{-1}$ ($lb_n \equiv k lb_n$), where 
$k=\sqrt E$ and the reflection phase time $\tau$ in units of inverse 
of incident energy $E$ ($\tau \equiv E\tau$).
After straight forward algebra in the large length ($L$) limit and in absence of magnetic flux, we obtain an analytical expression for the saturated phase delay time ( using Eq.~(\ref{phtm}) and Eq.~(\ref{ref}) ), which is given by,
\begin{equation}
\tau_s=\frac{\frac{1}{k\,\kappa_1}+\frac{k}{\kappa_1^3}}{\left(2\,+\,\frac{k^2}{2\kappa_1^2}\right)}\, .\label{satphtm}
\end{equation} 
Here the rectangular barrier has strength $V_1$ and width $lb_1=L$.
\begin{figure}[t]
\begin{center}
\includegraphics[width=8.0cm]{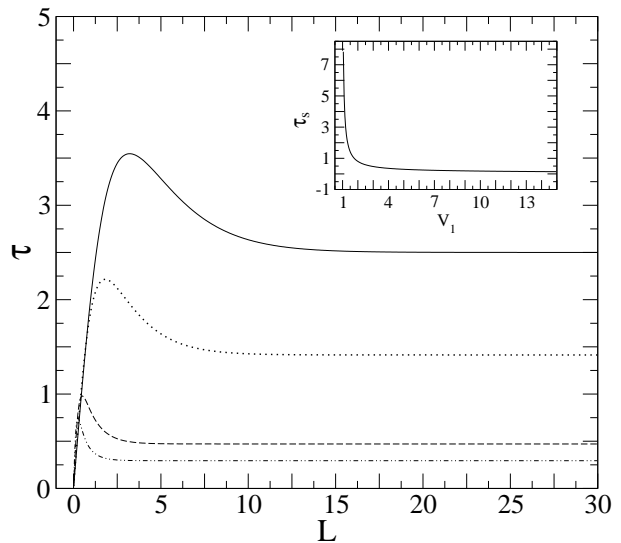}
\end{center}
\caption{In absence of magnetic flux ({\it i.e.} $\phi$ =0 ),  
for a ring with a barrier of strength $V_1$ throughout its circumference,
 the phase time $\tau$ is plotted as a function of ring's circumference $L$.
 The solid, dotted, dashed and dash-dotted curves are for $V_1 = 1.25, \, 1.5, \,3, \,5$ respectively. Incident energy is set to be $E=1$. In the inset the 
saturated value of phase time $\tau_s$ is plotted as a function of the barrier's
 strength for same $E$. 
} 
\label{fig:tau-L}
\end{figure}

First we take up a ring system with a single barrier along the circumference of the ring. For a tunneling particle having energy less than the barrier's strength we find out the reflection phase time $\tau$ as a function of barrier's width $L$ which in turn is the circumference of the ring. We see (Fig.~\ref{fig:tau-L}) that in absence of magnetic flux, $\tau$ evolves as a function of $L$ and asymptotically saturates to a value $\tau_s$ which is independent of $L$ thus confirming the `Hartman effect'. From Fig.~\ref{fig:tau-L} it is clear that the saturation value increases with the decreasing barrier-strength. In the inset of Fig.~\ref{fig:tau-L}, we have plotted $\tau_s$ as a function of barrier-strength. From this we can see that for electrons with incident energy close to the 
barrier-strength the value of $\tau_s$ is quite large. 

\begin{figure}[t]
\begin{center}
\includegraphics[width=8.0cm]{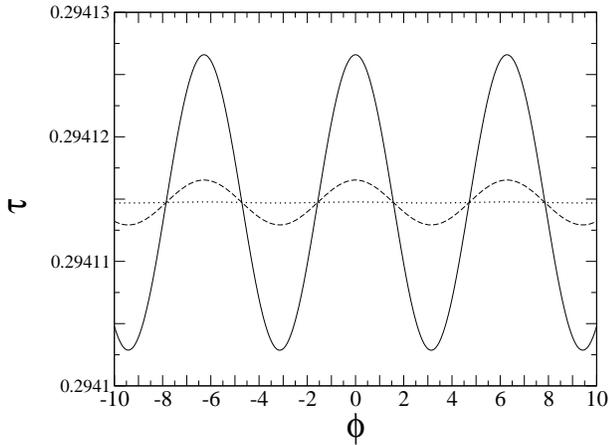}
\end{center}
\caption{For a ring with a barrier of strength 
$V_1$ lies throughout its circumference, the saturated phase time $\tau_s$ is 
plotted as a function of magnetic flux $\phi$. The solid, dashed and dotted 
curves are for $L = 6, 7, 9$ respectively. Other system parameters are $V_1=5$,
$E=1$.} 
\label{fig:tau-phi}
\end{figure}

To see the effect of magnetic flux on `Hartman effect', we consider the same
system but in presence of Aharonov-Bohm (AB) flux. We find out, for the tunneling particle, the reflection phase time as a function of embedded magnetic flux for different lengths $L$ of the barrier covering the ring's circumference. We have chosen the lengths such that in absence of the `AB-flux', for a given system ({\it i.e.} for known $E$ and $V$) the reflection phase time $\tau$ gets saturated in these lengths. From Fig.~\ref{fig:tau-phi} we see that  $\tau$ as a function
of $\phi$ shows AB-oscillations with an average value which is the saturation
value $\tau_s$ for the same system in absence of AB-flux. Further observe that (Fig.~\ref{fig:tau-phi}) $\tau$ is flux periodic with periodicity $\phi_0$. This is consistent with the fact that all the physical properties in presence of AB-flux across the ring must be periodic function of the flux with a period 
$\phi_0$ \cite{buttiker,webb,datta}. However, we see that as we increase $L$ the magnitude of AB-oscillation in $\tau$ decreases. Consequently in the large length limit the visibility vanishes. This clearly establishes `Hartman effect' even in presence of AB-flux. The constant value of $\tau$ thus obtained in the 
presence of flux is identical to $\tau_s$ in the absence of flux 
(see Fig.~(\ref{fig:tau-phi})) in the large length regime and its magnitude 
is given by Eq.~(\ref{satphtm}). This result clearly 
indicates that the reflection phase time in the presence of opaque barrier 
becomes not only independent of length of the circumference but also is 
independent of the AB-flux thereby observing the `Hartman effect' in the 
presence of AB-flux.  

\begin{figure}[t]
\begin{center}
\includegraphics[width=8.0cm]{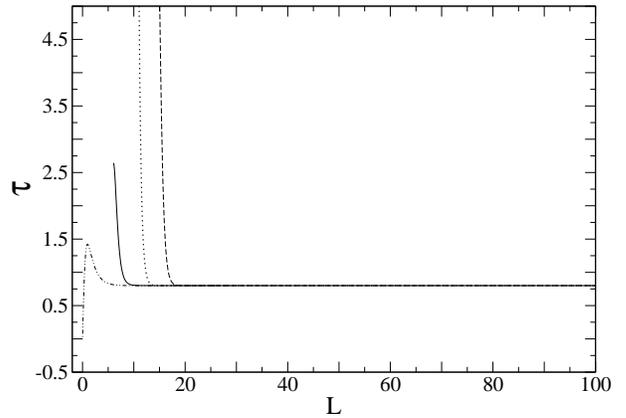}
\end{center}
\caption{In absence of magnetic flux ({\it i.e.} $\phi$ =0 ),  
for a ring with two barriers of strength $V_1$ and $V_3$ separated by an 
intermediate well region,
 the phase time $\tau$ is plotted as a function of ring's circumference $L$ 
for different width $w$ of the well. The dashed-dotted, solid, dotted and
dashed curves are for $w=0, 1, 5, 10$. Other system parameters are $lb_3=5$, 
$V_1=V_3=2$ and $E=1$.} 
\label{fig:tau-w}
\end{figure}

Now we consider the ring system with the ring having two successive barriers separated by an intermediate free space as shown in Fig.~\ref{system}. In absence 
of magnetic flux, we see the effect of `quantum well' on the reflection phase time $\tau$. In Fig.~\ref{fig:tau-w} $\tau$ is plotted as a function of one of the barrier's length (say $lb_1$) while other barrier's length is fixed ($lb_3=5$) 
and for few different values of length of the well. Here, the fixed value of the
barrier's length $lb_3$ is chosen in such a way that in absence of the well region the reflection phase time reaches saturation at this length. From Fig.~\ref{fig:tau-w} we see that for all parameter values of well's width, the saturation value of reflection phase time $\tau_s$ is same and it is equal to what we obtained in absence of the well in the ring system. Thus the saturated phase time becomes independent of the width of the well (in the long length limit)
in the off resonant case. This is as if the effective velocity of the  electron
in the well becomes infinite or equivalently length of the well
does not count (space collapse or space destroyer) while traversing the ring. 

\begin{figure}[t]
\begin{center}
\includegraphics[width=8.0cm]{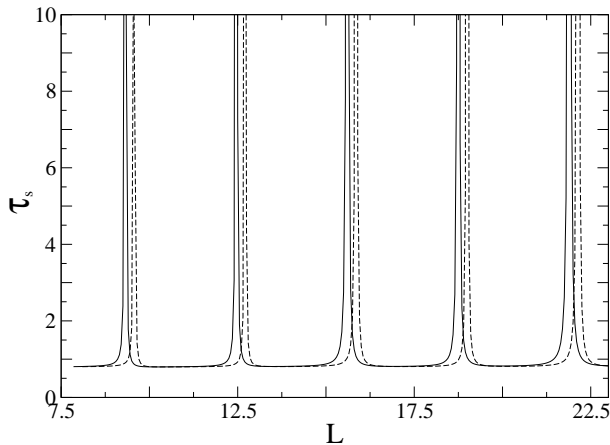}
\end{center}
\caption{In absence of magnetic flux ({\it i.e.} $\phi$ =0 ),  
for a ring with two barriers of strength $V_1$ and $V_3$ separated by an 
intermediate well region,
 the saturated phase time $\tau_s$ is plotted as a function of ring's circumference $L$ for different width $lb_1$ of the barrier. The solid and dashed curves are for $lb_1=2.75, 3$ respectively.
Other systems parameters are $lb_2=5$, $V_1=V_3=2$ and $E=1$.} 
\label{fig:resonance}
\end{figure}

Finally consider a similar system as that shown in Fig.~\ref{system}.
Here we see the effect of resonances, present in the ring system with a well, 
on the saturated reflection phase time $\tau_s$. For the system described above
with $V_1=V_3=2, lb_3=5, \phi=0$ we have plotted $\tau_s$, for the electrons with
incident energy $E=1$, as a function of the well's width for different parameter
values $lb_1$ in Fig.~\ref{fig:resonance}. We see that the resonances  which 
have Lorentzian shape become  sharper and  narrower as the width of the barrier
 $lb_1$ becomes large. For very large $lb_1$ the resonances are very hard to detect. 
It should be noted that as we increase the length of the well for fixed $E$ for
particular barrier lengths  incident energy $E$ coincides with resonances (or
resonant states) in the well (which arise due to constructive interference
due to multiple scatterings inside the well). For these values of lengths
we observe sharp rise in the saturated delay time and its magnitude
depends on the length of the well. It is worth to mention that away form
the resonance the value of $\tau_s$ is independent of the length of the
well (see Fig.\ref{fig:tau-w}) and depends only on the barrier strength.

\section{CONCLUSIONS}
We have generalized the Hartman effect beyond one dimension and
in the presence of magnetic flux. This is done by studying the
phase delay time in  a circular geometric ring in the reflection mode.
We have obtained an analytical expression for the saturated delay time.
We have also shown that in the presence of successive opaque barriers
separated by the potential well the saturated delay time becomes
independent of the width of the well (in the long length limit)
in the off resonant case. This is as if the effective velocity of the  electron
between the well becomes infinite or equivalently length
of the well
does not count (space collapse or space destroyer). We have further extended this effect
in  quantum mechanical networks with many arms  wherein  we have
shown additional existence of new  quantum non-local effects associated
with the phase delay time \cite{mynew}. Our reported results are amenable to experimental verifications for photon tunneling in appropriate  electromagnetic geometries.

\section{Acknowledgments}
One of the authors (SB) thanks Debasish Chaudhuri and Prof. Binayak Dutta-Roy
for several useful discussions.

\end{document}